\documentclass[graybox]{svmult}

\usepackage{mathptmx}       
\usepackage{helvet}         
\usepackage{courier}        
\usepackage{type1cm}        
\usepackage{makeidx}         
\usepackage{graphicx}        
\usepackage{multicol}        
\usepackage[bottom]{footmisc}

\makeindex             

\newcommand{\PEDIGREE}{{\sc PEDIGREE }}

\begin{document}

\title*{Modeling of pedestrians}
\author{Cecile Appert-Rolland}
\institute{Cecile Appert-Rolland \at Lab. of Theoretical Physics, CNRS UMR 8627, Univ. Paris-Sud, B\^at. 210, 91405 Orsay Cedex, France. \email{Cecile.Appert-Rolland@th.u-psud.fr}}
%
%
\maketitle

\abstract{Different families of models first developed
for fluid mechanics have been extended to road, pedestrian,
or intracellular transport.
These models allow to describe the systems at different
scales and to account for different aspects of dynamics.
In this paper, we focus on pedestrians and
illustrate the various families of models by
giving an example of each type.
We discuss the specificities of crowds compared
to other transport systems.
}

\section{Introduction}
\label{sec:intro}

What is the common point between fluids,
cars, pedestrians or molecular motors?
Though they are quite different and evolve
in systems of very different sizes,
they all result into flows, and they all
obey simple conservation laws.
As a result, the families of models
that have been developed in the past to describe
fluids at different scales have also been adapted
to describe highway traffic~\cite{chowdhury_s_s2000},
crowds~\cite{vicsek_z2012} or axonal transport~\cite{chou_m_z2011,appert-rolland_e_s2014}.

\begin{table}
\caption{Correspondence of model families, for four
different physical systems: fluids, road traffic,
pedestrian traffic and intracellular traffic.
We mention a few models (with their reference)
as prominent and/or historical examples
of a given model type.
The scale at which the system is described increases
as one goes down in the table.}
\label{tab1}       
\begin{tabular}{|p{2.8cm}|p{2.8cm}|p{2.8cm}|p{2.8cm}|}
\hline
&&&\\
Fluids & Road Traffic & Pedestrians &
\begin{minipage}{2.7cm}
Molecular Motors \end{minipage} \\
\hline
\hline
&&&\\
\textcolor{blue}{\small Molecular Dynamics} &
\textcolor{blue}{\small Car-following} &
\textcolor{blue}{\small Ped-following} &
\textcolor{blue}{\small Molecular Dynamics} \\
\cite{alder_w1959,rahman1964}
&
\cite{pipes1953,chandler_h_m1958,gazis_h_p1959}
&&\\
m {\bf a} = $\sum$ {\bf f} & a($\Delta V$, $\Delta x$) &
& m {\bf a} = $\sum$ {\bf f} \\
\hline
&&&\\
\textcolor{blue}{\small Kinetic theory} &
\textcolor{blue}{\small Kinetic theory} &
\textcolor{blue}{\small Kinetic theory} &
 \\
 P(v,x,t) & P(v,x,t) 
 & P(v,x,t,$\xi$)
  & \\
  \hline
  &&&\\
  \textcolor{blue}{\small Cellular automata} &
  \textcolor{blue}{\small Cellular automata} &
  \textcolor{blue}{\small Cellular automata} &
  \textcolor{blue}{\small Cellular automata} \\
&
  {\em \small Nagel-Schreckenberg} &
  {\em \small Floor Field}
&
{\em \small Langmuir} \\
 {\em \small FHP Model~}{\small \cite{frisch_h_p86}} & {\em model~}{\small \cite{nagel_s1992}} \textcolor{red}{} &
 {\em \small model~}{\small \cite{burstedde2001b}}
  & {\em \small kinetics~}{\small \cite{parmeggiani_f_f2004}} \textcolor{red}{} \\
  \hline
\textcolor{blue}{\small Continuous PDEs} &
\textcolor{blue}{\small Continuous PDEs} &
\textcolor{blue}{\small Continuous PDEs} & \textcolor{blue}{\small Continuous PDEs} \\
\begin{minipage}{2.7cm}
{\small  Conservation of mass
and momentum}
\end{minipage} &
\begin{minipage}{2.7cm}
{\small  Conservation of mass
$+$ fundamental diagram $j(\rho)$}
\end{minipage} &
\begin{minipage}{2.7cm}
{\small  Conservation of mass
and ...}
\end{minipage} &
\begin{minipage}{2.7cm}
{\small  Open system:
balance of fluxes}
\end{minipage}\\
  &&&\\
{\em \small Navier-Stokes Eqs} & {\em \small LWR Model~}{\small \cite{lighthill_w1955b,richards1956}} &
& \\
\hline
\end{tabular}
\end{table}

Let us consider first macroscopic models: 
At large scales, individuals are not visible
anymore, and the state of the system can
be characterized by locally averaged density
and velocity.
For fluids, Navier-Stokes equations express
the conservation of mass and of momentum.

For road traffic, mass conservation is still
relevant, and provides a first equation
relating density and velocity.
However, as vehicles are in contact
with the road, momentum is not conserved.
A second relation must be provided to close
the equations.
The simplest way is to give the
(possibly data-based) fundamental diagram,
relating the flow of vehicles and the density.
The resulting model is a so-called
first order model, a prominent example being the
LWR model~\cite{lighthill_w1955b,richards1956}.
The more sophisticated second-order
models~\cite{payne1971,aw_r2000}
express the fact that the adjustment of
flow to density may not be instantaneous
but rather takes place within a certain relaxation time.
The second relation between density and velocity
is then a second partial differential equation.

For pedestrians also, the mass conservation equation
must be completed to provide a closed set of equations.
However, the complexity is increased by the fact
that pedestrians, first, walk in a two-dimensional space,
and, second, do not necessarily all go in the same
direction.

Within cells, intracellular transport also involves
some ``walkers'', i.e. some molecules equipped with
some kind of legs that perform stepping along some cylindrical
tracks called microtubules.
In contrast with human pedestrians,
these so-called molecular motors do not only walk
along microtubules, they can also detach
from the microtubules, diffuse around, and attach again.
Thus, if one considers the density of motors on the
microtubule, even mass conservation is not realized
any more.
The equations that determine the evolution
of density and velocity must then rather express
some balance of fluxes between different regions of the
system.

In the same way as various macroscopic models
can be proposed for all these systems, there
are some equivalents of molecular dynamics
or of cellular automata approaches that have
been developed for road, pedestrian or
intracellular traffic.

In most cases, for a given physical system,
different types of models have been proposed
independently to account for the behavior of the
system at different scales, leading to large
families of models.
In some cases however, it is possible to relate the models
at the different scales and to understand how
the macroscopic behavior can emerge from the individual
dynamics.

In this paper, we shall focus on pedestrian modeling,
and give an example for each family of models.
Part of this work (sections \ref{sec:pedfol}, \ref{sec:macro},
and part of \ref{sec:kin})
was performed in the frame of
the interdisciplinary \PEDIGREE project~\cite{pedigree}.
The teams involved are presented in Table~\ref{tab2}.

The work of section \ref{sec:ca} was performed as part
of the master and PhD of Julien Cividini,
in collaboration with H. Hilhorst.

\begin{table}
\caption{The \PEDIGREE Project involved four French teams listed
below.
}
\label{tab2}       
\begin{tabular}{|p{2.0cm}|p{2.0cm}|p{2.0cm}|p{2.0cm}|p{2.4cm}|}
\hline
\vskip 0.01cm &&&&\\
Laboratory & IMT & INRIA & CRCA & LPT \\
\hline
\vskip 0.01cm &&&&\\
\begin{minipage}{1.9cm}
Team Leader \end{minipage} & P. Degond & J. Pettr\'e & G. Theraulaz & C. Appert-Rolland \\
\hline
\vskip 0.01cm &&&&\\
Participants & J. Fehrenbach & S. Donikian & O. Chabiron & J. Cividini \\
 & J. Hua & S. Lemercier & E. Guillot & A. Jeli\'c \\
 & S. Motsch & & M. Moreau & \\
 & J. Narski & & M. Moussa{\"i}d & \\
\hline
\end{tabular}
\end{table}

\section{Ped-following model}
\label{sec:pedfol}

Fluids can be described at the level of molecules,
by taking into account all the interaction potentials
between atoms in a more or less refined way, as is done
in molecular dynamics simulations~\cite{alder_w1959,rahman1964}.
When vehicles or pedestrians are considered,
two main difficulties arise. First the interaction potential
is not know - actually the interaction cannot in general
be written as deriving from a potential. Second,
the interaction is in general highly non-isotropic,
and does not depend only on the position but also
on the velocity and on the target direction of each
individual.

In road traffic, cars naturally follow lanes.
This features greatly simplifies the problem.
Each car has a single well-defined predecessor on its lane.
Apart from lane changes, a car driver can only
adjust its speed. He will do so depending on the conditions
in front
(distance, velocity, acceleration of the predecessor).
Actually several cars ahead could be taken into account
(and indeed some empirical
studies~\cite{hoogendoorn_o_s2006b}
have shown that a driver
may take into account several of its leaders). But still,
there is a clear hierarchy among the leaders, given
by their order in the lane.

In pedestrian traffic, individuals evolve in a
two-dimensional space, and may interact with several
pedestrians at the same time, without a clear hierarchy.
Besides, the combination of interactions is in general
not a simple sum of one-by-one interactions.
However, there are situations where the flow is organized
in such a way that it is quasi one dimensional.

For example in corridors, all pedestrians mostly go
in the same direction. Even if two opposite flows
are considered, it is known that some lanes are
formed spontaneously, and within each lane
the flow is again quasi one dimensional and one directional.

The way pedestrians follow each other is even more clear when
pedestrians
walk in a line.
Such a configuration can be met for example
in very narrow corridors.
It has been realized in several experiments~\cite{seyfried2005,chattaraj_s_c2009a,yanagisawa_t_n2012a}, in order
to study how pedestrians react when they can only
adjust their speed.
One may then wonder how the acceleration of a pedestrian
is related to the distance, velocity, acceleration of
its predecessor, and how the behavior of a pedestrian
differs from the one of a car.
However, in order to evaluate the following behavior of
a pedestrian, one needs to be able to track at the
same time, and on long enough time windows, the
trajectory of both the pedestrian under consideration
and its predecessor.

Such an experiment has been realized in
the frame of the \PEDIGREE project~\cite{pedigree}.
Pedestrians were asked to walk as a line, i.e.
to follow each other without passing~\cite{lemercier2012a}.
Their trajectory was circular, in order to avoid
boundary effects.
The motion of all pedestrians was tracked with
a high precision motion capture device (VICON)~\cite{exp-m2s}.
As a result, the trajectories of all pedestrians
were obtained for the whole duration of the
experiment (from 1 to 3 minutes).

Various combinations of the dynamic coordinates
of the predecessor have been tested against
the acceleration $a$ of the follower.
It turned out that the best correlation was
obtained~\cite{lemercier2012a,appert-rolland2012} for the relation
\begin{equation}
a(t) = C \frac{\Delta v(t-\tau)}{\left[\Delta x(t)\right]^\gamma}
\end{equation}
where $v$ is the velocity of the predecessor, and
$\Delta x$ the distance between the predecessor
and its follower.

One important difference with car traffic is the
time delay $\tau$ introduced in the velocity:
While the follower is able to evaluate quite 
instantaneously the position of his predecessor,
he needs some time delay $\tau$ to evaluate his velocity.

Another difference with car traffic is the ability of
pedestrians to flow even at very large local densities~\cite{jelic2012a}.
In the aforementioned experiment, the velocity was
still of the order of $1$ or $2$ dm/s at local densities as
high as $3$ ped/m.
This can be achieved thanks to the ability of pedestrians to keep
walking even at very low densities: they can reduce the amplitude
of their steps almost to zero while still keeping a stepping
pace almost constant~\cite{jelic2012b}.

In contrast to cars, pedestrians can also take advantage of any
space left by the predecessor, synchronizing partially their
steps as was
observed in previous experiments~\cite{seyfried2005}.
Surprisingly, this synchronization effect is also observed
for pedestrians walking at a larger distance~\cite{jelic2012b},
probably as a result of the tendency of pedestrians to synchronize
with external rhythmic stimuli~\cite{yanagisawa_t_n2012a}.

Here we have presented a model for one-dimensional pedestrian
flows. In general, pedestrians move in a two-dimensional space,
and various agent based models have been proposed which
we shall not review here.

\section{One-dimensional bi-directional macroscopic model for crowds}
\label{sec:macro}

At the other extreme, when seen from a distance, crowds
can be described as continuous fluids.
As mentioned in the introduction, one important difference
with fluids is that pedestrians have a target - which
may not be the same for all of them.
A simple configuration is met in corridors:
the flow is quasi one-dimensional, but pedestrians can
walk in both directions.
There is thus a need to distinguish two densities $\rho_\pm$ of
pedestrians, one for each walking direction.
Each density obeys a conservation law:
\begin{eqnarray*}
    & & \hspace{-1cm} \partial_t \rho_+ + \partial_x (\rho_+ u_+) = 0 , \label{2AR_n+} \\
    & & \hspace{-1cm} \partial_t \rho_- + \partial_x (\rho_- u_-) = 0 , \label{2AR_n-} 
\end{eqnarray*}
where $u_\pm$ is the locally averaged velocity of pedestrians going in the $\pm$ direction.

Two other relations are needed to determine the four unknown densities and velocities.
This is achieved by writing two other differential equations for the
momentum~\cite{appert-rolland_d_m2011,appert-rolland_d_m2012}
\begin{eqnarray*}
  & & \hspace{-1cm} \partial_t (\rho_+ u_+)  + \partial_x (\rho_+ u_+ u_+ ) =  - \rho_+ \,
  \left( \frac{d}{dt} \right)_+ [ p(\rho_+,\rho_-) ]  , \label{2AR_rho_u+} \\
  & & \hspace{-1cm} \partial_t (\rho_- u_-)  + \partial_x (\rho_- u_- u_- ) =   \rho_- \,
  \left( \frac{d}{dt} \right)_- [ p(\rho_-,\rho_+) ]  , \label{2AR_rho_u-}
\end{eqnarray*}
in which, by analogy to the pressure in fluid mechanics, the interactions between pedestrians
are described by a term $p(\rho_\pm,\rho_\mp)$.
There is however a major difference with fluid mechanics:
following~\cite{aw_r2000}, 
the derivative 
\begin{equation}
(d/dt)_{\pm} = \partial_t + u_{\pm} \partial_x
\end{equation}
is taken in the referential of the walking pedestrians, and not in the fixed frame as for fluids.
Indeed, pedestrians react to their perception of the surrounding
density as they see it while walking.

The term $p(\rho_\pm,\rho_\mp)$ is actually not a pressure
as in fluid mechanics, but rather a velocity offset between
the achieved velocity $u_\pm$, and another quantity $w_\pm$
which, as it is conserved along each pedestrian trajectory,
can be interpreted as the desired velocity that the pedestrian
would have if he was alone.
In other words,
\begin{eqnarray*}
u_+ & = & w_+ - p(\rho_+, \rho_-)\\
- u_- & = & w_- - p(\rho_-, \rho_+)
\end{eqnarray*}
where $w_\pm$ are Riemann invariants
\begin{eqnarray*}
\partial_t w_+  + u_+ \partial_x w_+ & = & 0 \\
\partial_t w_-  + u_- \partial_x w_- & = & 0
\end{eqnarray*}
conserved along the trajectories of $\pm$ pedestrians.
The function $p(\rho_\pm,\rho_\mp)$ can be determined
from experimental measurements.

One difficulty coming from the fact that pedestrians
do not necessarily have the same target site is that
they may converge towards the same region, leading
in the simulations to density divergences.
A special treatment is thus required in order
to limit the density to physical values.
The solution retained in~\cite{appert-rolland_d_m2011,appert-rolland_d_m2012}
was to let $p(\rho_\pm,\rho_\mp)$ diverge when the density
approaches its upper limit value.

\section{Kinetic models}
\label{sec:kin}

In kinetic models, instead of describing explicitely
the presence of a particle (molecule, vehicle or pedestrian) at a given
location with a given velocity, one deals with the corresponding probability.

\begin{figure}[b]
\includegraphics[width=\linewidth]{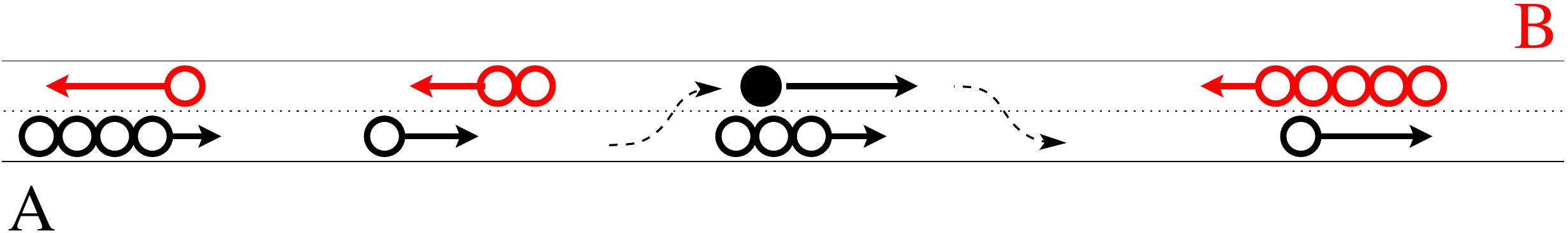}
\caption{Kinetic model for a bidirectional two-lane road.
The vehicles (red and black circles) have different desired velocities,
leading to platoon formation. Overtaking can take place only
if there is enough free space on the other lane. From~\cite{appert-rolland_h_s2010}.}
\label{fig:kinetic}       
\end{figure}

For example, in~\cite{appert-rolland_h_s2010}, with G. Schehr and H. Hilhorst,
we have considered the case of a bidirectional two-lane road.
Cars had different desired velocities, but
the overtaking can occur only if there is enough space on the other lane
(see Fig.~\ref{fig:kinetic}).
Assuming translation invariance of the probability distributions
along the road, the problem results in finding the distribution
of effective velocities as a function of the distribution of
desired velocities.
The solution of this problem requires to evaluate the overtaking probability.
To do so, we assume that
the probability to find a sufficient empty interval in the
other lane at a given time and given place is equal
to the average probability (mean-field assumption).
Under these assumptions, we find that a symmetry breaking can
occur between the lanes.
This model developed for road traffic could be seen
as a first attempt to model
pedestrians in a corridor when lanes are formed, i.e.
at high enough densities. However, this organization
into lanes will not be as stable as in road traffic.

To account more completely for pedestrian flows, one has to consider
the full joint probability distribution $f({\bf v},{\bf x},t,{\bf \xi})$
of finding in position {\bf x} at time $t$ a pedestrian with
velocity {\bf v} and target site located in {\bf $\xi$}.
It is out of scope yet to find universal equations for this distribution.
However, one may try to derive those from microscopic models.
Two such derivations have been proposed by P. Degond et al
in \cite{degond2013a} and \cite{degond2013b}, starting from
agent based models in which pedestrians modify their direction
(and possibly velocity modulus) in order to avoid
possible collisions in the near future, while still trying 
to keep as close as possible to their target direction.
Some mean-field approximations have to be done to go from
the microscopic discrete models to the kinetic ones.
The smoothing due to these mean-field expressions has to be
balanced by the introduction of some appropriate noise in the 
equations for the probability distribution $f$.

Once they have been obtained, these kinetic models can
themselves be taken as a starting point to derive
macroscopic models~\cite{degond2013a,degond2013b}
in two dimensions.

\section{Cellular automata}
\label{sec:ca}

To complete our comparison between pedestrian and fluid models,
we shall now consider cellular automata models.
For fluids, the story started in 1986, with the FHP model~\cite{frisch_h_p86},
in which pointlike particles were hopping onto a hexagonal lattice,
and undergoing collisions at the nodes of the lattice.
Providing that these collisions conserve mass and momentum
but still mix enough the particle distributions,
and that the lattice has enough symmetries,
the resulting lattice gas was found to obey equations very
close to Navier-Stokes equations~\cite{gouyet_a02}. Hence direct simulations
of this lattice gas were providing solutions of the (almost) Navier-Stokes
equations - a breakthrough given the difficulty to solve the latter.

A similar approach was proposed in 1992 for road traffic by
Nagel and Schreckenberg~\cite{nagel_s1992}.
Of course in this case there is no momentum conservation anymore,
but rather some rules expressing the increase of velocity
up to some maximal velocity, under the constraint of collision avoidance.

For pedestrians, interactions can be quite long ranged and one has
to combine interactions taking place in any direction.
It is thus a priori quite complex to develop a cellular
automaton based on interactions between neighboring cells.
A solution inspired from ants was provided by the use of some effective pheromones,
which mediate the interactions between
pedestrians~\cite{schadschneider_k_n2003,burstedde2001b,nishinari2006}.

Apart from the geometry of the lattice and the evolution rules,
a cellular automaton is also defined by the sequence under which lattice sites are
updated.
Processes in continuous time, with independent events occurring with given
rates, are well described by random sequential updates, in which a site
is chosen at random and updated at each micro-time step.
However this update leads to large fluctuations (the same site can be chosen
twice in a row while another one will be ignored for a long time).
Thus, for traffic applications, more regular updates are preferred.
In particular the parallel update in which all the sites are updated
in parallel at discrete time steps ensures a certain regularity in the flow.
It introduces a time scale (the aforementioned time step) which can
be interpreted as the reaction time of individuals.

Parallel update is widely used in road traffic modeling.
It is also employed for pedestrians~\cite{kirchner2003,kirchner_n_s2003},
but requires to be complemented
by extra rules. Indeed, in two-dimensional flows, two pedestrians
may chose the same target site, resulting in a conflict that has to be solved
by ad-hoc rules. Though these conflicts may be given a physical meaning
in terms of friction~\cite{kirchner_n_s2003}, some other updates which
do not require these extra rules have been proposed.

The random shuffle update~\cite{wolki_s_s2006,wolki_s_s2007b,smith_w2007a,klupfel2007a}
ensures that each site will be updated
exactly once per time step, but in an order that is randomly chosen
at each time step.
This update has indeed been used in early cellular automata
simulations of pedestrian
evacuations~\cite{klupfel2000,klupfel2007a}\footnote{Note that
in some communities, {\em random shuffle update} is called {\em
random sequential update}, as done in~\cite{klupfel2000}. We
shall stick to the denomination used in physics, for which {\em
random sequential update} rather refers to an update close to
continuous time.}.

The frozen shuffle update~\cite{appert-rolland_c_h2011a,appert-rolland_c_h2011b}
associates to each pedestrian a fixed phase $\tau$, i.e.
a real number between $0$ and $1$, and updates pedestrians once per time step
in the order of increasing phases.
This update allows to have higher fluxes than parallel update,
reproducing the tendency of pedestrians to flow even
at high densities.
Besides, the phase $\tau$ can be given a physical meaning.
It can represent the phase in the stepping cycle.
It allows also to some extend to map the cellular automaton
dynamics onto a continuous time/continuous space
dynamics~\cite{appert-rolland_c_h2011b}.

Cellular automata simulations can be useful to simulate
large systems~\cite{klupfel2007a}.
They can also help to understand some pattern formation~\cite{hoogendoorn_d2003},
for example lane formation in
counterflows~\cite{burstedde2001a,burstedde2001b}, or diagonal
patterns at the crossing of two perpendicular
flows~\cite{cividini_a_h2013,cividini2014}.

\section{Conclusion}

In this paper we have reviewed a few models for pedestrians
proposed in the past years, to
illustrate the various families of models that
span over the different physical systems considered
at the TGF conference.

For completeness, we must mention that at even larger scales
than considered in this paper,
models for road or pedestrian traffic must be supplemented with
route choice models, as in~\cite{hoogendoorn_b2004b}.

\begin{acknowledgement}
The \PEDIGREE project has been supported by the French 'Agence Nationale pour la Recherche (ANR)' (contract
number ANR-08-SYSC-015-01, from 2008 to 2011).

Subsequent data analysis was partially supported
by the `RTRA Triangle de la physique' (Project 2011-033T).

\end{acknowledgement}

\bibliographystyle{elsart-num}

\begin{thebibliography}{10}

\bibitem{chowdhury_s_s2000}
D.~Chowdhury, L.~Santen, A.~Schadschneider, Statistical physics of vehicular
  traffic and some related systems, Phys. Reports 329 (2000) 199.

\bibitem{vicsek_z2012}
T.~Vicsek, A.~Zafeiris, Collective motion, Physics Reports 517~(3-4) (2012)
  71--140.

\bibitem{chou_m_z2011}
T.~Chou, K.~M. R. K.~P. Zia, Non-equilibrium statistical mechanics: from a
  paradigmatic model to biological transport, Reports on progress in physics 74
  (2011) 116601.

\bibitem{appert-rolland_e_s2014}
C.~Appert-Rolland, M.~Ebbinghaus, L.~Santen, Intracellular transport driven by
  cytoskeletal motors~: General mechanisms and defects, submitted to Phys. Rep.

\bibitem{alder_w1959}
B.~Alder, T.~E. Wainwright, Studies in molecular dynamics. {I}. general method,
  J. Chem. Phys. 31 (1959) 459.

\bibitem{rahman1964}
A.~Rahman, Correlations in the motion of atoms in liquid argon, Phys. Rev. 136
  (1964) A405A411.

\bibitem{pipes1953}
L.~Pipes, An operational analysis of traffic dynamics, J. of Applied Physics 24
  (1953) 274--281.

\bibitem{chandler_h_m1958}
R.~Chandler, R.~Herman, E.~Montroll, Traffic dynamics: Studies in car
  following, Operations Research 6 (1958) 165--184.

\bibitem{gazis_h_p1959}
D.~Gazis, R.~Herman, R.~Potts, Car following theory of steady state traffic
  flow, Operations Research 7 (1959) 499--505.

\bibitem{frisch_h_p86}
U.~Frisch, B.~Hasslacher, Y.~Pomeau, Lattice-gas automata for the
  {N}avier-{S}tokes equation, Phys. Rev. Lett. 56 (1986) 1505--1508.

\bibitem{nagel_s1992}
K.~Nagel, M.~Schreckenberg, A cellular automaton model for freeway traffic, J.
  Phys. I 2 (1992) 2221--2229.

\bibitem{burstedde2001b}
C.~Burstedde, K.~Klauck, A.~Schadschneider, J.~Zittartz, Simulation of
  pedestrian dynamics using a 2-dimensional cellular automaton, Physica A 295
  (2001) 507--525.

\bibitem{parmeggiani_f_f2004}
A.~Parmeggiani, T.~Franosch, E.~Frey, Totally asymmetric simple exclusion
  process with {L}angmuir kinetics, Phys. Rev. E 70 (2004) 046101.

\bibitem{lighthill_w1955b}
M.~Lighthill, G.~Whitham, On kinematic waves. {II}. a theory of traffic flow on
  long crowded roads, Proceedings of the Royal Society of London. Series A,
  Mathematical and Physical Sciences A 229 (1955) 317--345.

\bibitem{richards1956}
P.~Richards, Shock waves on the highway, Operations research 4 (1956) 42--51.

\bibitem{payne1971}
H.~Payne, Models of freeway traffic and control, Simulation Councils Proc.
  Series: Mathematical Model of Public Systems 1 (1971) 51--60.

\bibitem{aw_r2000}
M.~R. A.~Aw, Resurrection of ``second order'' models of traffic flow and
  numerical simulation, SIAM Journal on Applied Mathematics 60 (2000) 916--938.

\bibitem{hoogendoorn_o_s2006b}
S.~Hoogendoorn, S.~Ossen, M.~Schreuder, Empirics of multianticipative
  car-following behavior, Transportation Research Record 1965 (2006) 112--120.

\bibitem{seyfried2005}
A.~Seyfried, B.~Steffen, W.~Klingsch, M.~Boltes, The fundamental diagram of
  pedestrian movement revisited, J. Stat. Mech. (2005) P10002.

\bibitem{chattaraj_s_c2009a}
U.~Chattaraj, A.~Seyfried, P.~Chakroborty, Comparison of pedestrian fundamental
  diagram across cultures, Advances in Complex Systems 12 (2009) 393--405.

\bibitem{yanagisawa_t_n2012a}
D.~Yanagisawa, A.~Tomoeda, K.~Nishinari, Improvement of pedestrian flow by slow
  rhythm, Phys. Rev. E 85 (2012) 016111.

\bibitem{lemercier2012a}
S.~Lemercier, A.~Jelic, R.~Kulpa, J.~Hua, J.~Fehrenbach, P.~Degond,
  C.~Appert-Rolland, S.~Donikian, J.~Pettr\'e, Realistic following behaviors
  for crowd simulation, COMPUTER GRAPHICS FORUM 31 (2012) 489--498.

\bibitem{appert-rolland2012}
C.~Appert-Rolland, A.~Jelic, P.~Degond, J.~Fehrenbach, J.~Hua, A.~Cr\'etual,
  R.~Kulpa, A.~Marin, A.-H. Olivier, S.~Lemercier, J.~Pettr\'e, Experimental
  study of the following dynamics of pedestrians, in: U.~Weidmann, U.~Kirsch,
  M.~S. (Eds.), Pedestrian and Evacuation Dynamics 2012, Springer,
  Heidelberg, 2014, pp. 305--316.

\bibitem{jelic2012a}
A.~Jeli\'c, C.~Appert-Rolland, S.~Lemercier, J.~Pettr\'e, Properties of
  pedestrians walking in line -- fundamental diagrams, Phys. Rev. E 85 (2012)
  036111.

\bibitem{jelic2012b}
A.~Jeli\'c, C.~Appert-Rolland, S.~Lemercier, J.~Pettr\'e, Properties of
  pedestrians walking in line. ii. stepping behavior, Phys. Rev. E 86 (2012)
  046111.

\bibitem{appert-rolland_d_m2011}
C.~Appert-Rolland, P.~Degond, S.~Motsch, Two-way multi-lane traffic model for
  pedestrians in corridors, Networks and Heterogeneous Media 6 (2011) 351--381.

\bibitem{appert-rolland_d_m2012}
C.~Appert-Rolland, P.~Degond, S.~Motsch, A macroscopic model for bidirectional
  pedestrian flow, in: U.~Weidmann, U.~Kirsch, M.~S. (Eds.), Pedestrian
  and Evacuation Dynamics 2012, Springer, Heidelberg, 2014, pp. 575--584.

\bibitem{appert-rolland_h_s2010}
C.~Appert-Rolland, H.~Hilhorst, G.~Schehr, Spontaneous symmetry breaking in a
  two-lane model for bidirectional overtaking traffic, J. Stat. Mech. (2010)
  P08024.

\bibitem{degond2013a}
P.~Degond, C.~Appert-Rolland, M.~Moussaid, J.~Pettr\'e, G.~Theraulaz, A
  hierarchy of heuristic-based models of crowd dynamics, J. Stat. Phys. 152
  (2013) 1033--1068.

\bibitem{degond2013b}
P.~Degond, C.~Appert-Rolland, J.~Pettr\'e, G.~Theraulaz, Vision-based
  macroscopic pedestrian models, Kinetic and Related Models 6 (2013) 809--839.

\bibitem{gouyet_a02}
J.-F. Gouyet, C.~Appert, Stochastic and hydrodynamic lattice gas models:
  mean-field kinetic approaches, Int. J. Bifurcat. Chaos 12 (2002) 227--259.

\bibitem{schadschneider_k_n2003}
A.~Schadschneider, A.~Kirchner, K.~Nishinari, From ant trails to pedestrian
  dynamics, Applied Bionics and Biomechanics 1 (2003) 11--19.

\bibitem{nishinari2006}
K.~Nishinari, K.~Sugawara, T.~Kazama, A.~Schadschneider, D.~Chowdhury,
  Modelling of self-driven particles: Foraging ants and pedestrians, Physica A
  372 (2006) 132--141.

\bibitem{kirchner2003}
A.~Kirchner, H.~Kl{\"u}pfel, K.~Nishinari, A.~Schadschneider, M.~Schreckenberg,
  Simulation of competitive egress behaviour: comparison with aircraft
  evacuation data, Physica A 324 (2003) 689--697.

\bibitem{kirchner_n_s2003}
A.~Kirchner, K.~Nishinari, A.~Schadschneider, Friction effects and clogging in
  a cellular automaton model for pedestrian dynamics, Phys. Rev. E 67 (2003)
  056122.

\bibitem{wolki_s_s2006}
M.~W{\"o}lki, A.~Schadschneider, M.~Schreckenberg, Asymmetric exclusion
  processes with shuffled dynamics, J. Phys. A-Math. Gen. 39 (2006) 33--44.

\bibitem{wolki_s_s2007b}
M.~W{\"o}lki, M.~Schadschneider, M.~Schreckenberg, Fundamental diagram of a
  one-dimensional cellular automaton model for pedestrian flow -- the {ASEP}
  with shuffled update, in: ed.: N.~Waldau, P.~Gattermann, H.~Knoflacher,
  M.~Schreckenberg (Eds.), Pedestrian and Evacuation Dynamics 2005, Berlin,
  Springer, 2007, p. 423.

\bibitem{smith_w2007a}
D.~A. Smith, R.~E. Wilson, Dynamical pair approximation for cellular automata
  with shuffle update, J. Phys. A: Math. Theor. 40~(11) (2007) 2651--2664.

\bibitem{klupfel2007a}
H.~Kl{\"u}pfel, The simulation of crowds at very large events, in:
  A.~Schadschneider, T.~Poschel, R.~Kuhne, M.~Schreckenberg, D.~Wolf (Eds.),
  Traffic and Granular Flow ' 05, 2007, pp. 341--346.

\bibitem{klupfel2000}
H.~Kl{\"u}pfel, T.~Meyer-K{\"o}nig, J.~Wahle, M.~Schreckenberg, Microscopic
  simulation of evacuation processes on passenger ships, in: S.~Bandini,
  T.~Worsch (Eds.), Proceedings of the 4th International Conference on Cellular
  Automata for Research and Industry (ACRI00), Springer, 2000, pp. 63--71.

\bibitem{appert-rolland_c_h2011a}
C.~Appert-Rolland, J.~Cividini, H.~Hilhorst, Frozen shuffle update for an
  asymmetric exclusion process on a ring, J. Stat. Mech. (2011) P07009.

\bibitem{appert-rolland_c_h2011b}
C.~Appert-Rolland, J.~Cividini, H.~Hilhorst, Frozen shuffle update for a
  deterministic totally asymmetric simple exclusion process with open
  boundaries, J. Stat. Mech. (2011) P10013.

\bibitem{hoogendoorn_d2003}
S.~P. Hoogendoorn, W.~Daamen, Self-organization in walker experiments, in:
  S.~Hoogendoorn, S.~Luding, P.~Bovy, et~al. (Eds.), Traffic and Granular Flow
  '03, Springer, 2005, p.~??

\bibitem{burstedde2001a}
C.~Burstedde, A.~Kirchner, K.~Klauck, A.~Schadschneider, J.~Zittartz, Cellular
  automaton approach to pedestrian dynamics - applications, in:
  M.~Schreckenberg, S.~S. (Eds.), Pedestrian and Evacuation Dynamics,
  Springer, 2001, p.~87.

\bibitem{cividini_a_h2013}
J.~Cividini, C.~Appert-Rolland, H.~Hilhorst, Diagonal patterns and chevron
  effect in intersecting traffic flows, Europhys. Lett. 102 (2013) 20002.

\bibitem{cividini2014}
J.~Cividini, Generic instability at the crossing of pedestrian flows, in:
  Traffic and Granular Flow '13, Springer, 2014.

\bibitem{hoogendoorn_b2004b}
S.~P. Hoogendoorn, P.~H.~L. Bovy, Pedestrian route-choice and activity
  scheduling theory and models, Transportation Research Part B: Methodological
  38 (2004) 169--190.

\bibitem{pedigree}
\PEDIGREE project: website http://www.math.univ-toulouse.fr/pedigree

\bibitem{exp-m2s}
Experiments were
organized and realized by the \PEDIGREE partnership at
University Rennes 1, with the help of the laboratory M2S
from Rennes 2.

\end{thebibliography}

\end{document}